\documentclass[sigconf]{acmart}

\AtBeginDocument{%
  \providecommand\BibTeX{{%
    \normalfont B\kern-0.5em{\scshape i\kern-0.25em b}\kern-0.8em\TeX}}}

\usepackage[hide]{chato-notes}

\copyrightyear{2024}
\acmYear{2024}
\setcopyright{rightsretained}
\acmConference[FRAME '24]{Workshop on Flexible Resource and Application Management on the Edge}{June 3--7, 2024}{Pisa, Italy}
\acmBooktitle{Workshop on Flexible Resource and Application Management on the Edge (FRAME '24), June 3--7, 2024, Pisa, Italy. {\bf This is the author's version of the work. It is posted here for your personal use. Not for redistribution.} The definitive Version of Record was published by}\acmDOI{10.1145/3659994.3660315}
\acmISBN{979-8-4007-0641-7/24/06}

\begin{document}

\title{Urgent Edge Computing}



\author{Patrizio Dazzi}
\orcid{1234-5678-9012}
\affiliation{%
  \institution{Department of Computer Science \\ University of Pisa}
  \streetaddress{Largo Bruno Potecorvo, 3}
  \city{Pisa}
  \state{PI}
  \country{Italy}
  \postcode{56127}
}
\email{patrizio.dazzi@unipi.it}

\author{Luca Ferrucci}
\orcid{1234-5678-9012}
\affiliation{%
  \institution{Department of Computer Science \\ University of Pisa}
  \streetaddress{Largo Bruno Potecorvo, 3}
  \city{Pisa}
  \state{PI}
  \country{Italy}
  \postcode{56127}
}
\email{luca.ferrucci@unipi.it}

\author{Marco Danelutto}
\orcid{1234-5678-9012}
\affiliation{%
  \institution{Department of Computer Science \\ University of Pisa}
  \streetaddress{Largo Bruno Potecorvo, 3}
  \city{Pisa}
  \state{PI}
  \country{Italy}
  \postcode{56127}
}
\email{marco.danelutto@unipi.it}

\author{Konstantinos Tserpes}
\affiliation{%
  \institution{Department of Informatics and Telematics, Harokopio University of Athens and School of Electrical and Computer Engineering, National Technical University of Athens}
  \city{Athens}
  \state{Attika}
  \country{Greece}
  \postcode{56127}}
\email{tserpes@hua.gr}

\author{Antonis Makris}
\affiliation{%
  \institution{Department of Informatics and Telematics, Harokopio University of Athens and School of Electrical and Computer Engineering, National Technical University of Athens}
  \city{Athens}
  \state{Attika}
  \country{Greece}
  \postcode{}}
\email{amakris@hua.gr}

\author{Theodoros Theodoropoulos}
\affiliation{%
  \institution{Department of Informatics and Telematics, Harokopio University of Athens and School of Electrical and Computer Engineering, National Technical University of Athens}
  \city{Athens}
  \state{Attika}
  \country{Greece}
  \postcode{}}
\email{ttheod@hua.gr}

\author{Jacopo Massa}
\orcid{1234-5678-9012}
\affiliation{%
  \institution{Department of Computer Science \\ University of Pisa and Institute of Information Science and Technologies \\ National Research Council of Italy}
  \streetaddress{Largo Bruno Potecorvo, 3}
  \city{Pisa}
  \state{PI}
  \country{Italy}
  \postcode{56127}
}
\email{jacopo.massa@phd.unipi.it}

\author{Emanuele Carlini}
\affiliation{%
  \institution{Institute of Information Science and Technologies \\ National Research Council of Italy}
  \city{Pisa}
  \state{PI}
  \country{Italy}
  \postcode{56127}}
\email{emanuele.carlini@isti.cnr.it}

\author{Matteo Mordacchini}
\affiliation{%
  \institution{Institute of Informatics and Telematics \\ National Research Council of Italy}
  \city{Pisa}
  \state{PI}
  \country{Italy}
  \postcode{56127}}
\email{matteo.mordacchini@iit.cnr.it}

\renewcommand{\shortauthors}{Dazzi et al.}
\newcommand{\UEC}{Urgent Edge Computing\xspace}
\newcommand{\UC}{Urgent Computing\xspace}

\begin{abstract}
This position paper introduces \UEC (UEC) as a paradigm shift addressing the evolving demands of time-sensitive applications in distributed edge environments, in time-critical scenarios. With a focus on ultra-low latency, availability, resource management, decentralization, self-organization, and robust security, UEC aims to facilitate operations in critical scenarios such as disaster response, environmental monitoring, and smart city management. This paper outlines and discusses the key requirements, challenges, and enablers along with a conceptual architecture. The paper also outlines the potential applications of \UEC.
\end{abstract}

\ccsdesc[300]{Computer systems organization~Dependable and fault-tolerant systems and networks}
\ccsdesc[300]{Human-centered computing~Ubiquitous and mobile computing theory, concepts and paradigms}
\ccsdesc[300]{Networks~Overlay and other logical network structures}
\ccsdesc[300]{Networks~Location based services}
\ccsdesc[300]{Information systems~Information systems applications}
\ccsdesc[500]{Computer systems organization~Distributed architectures}

\keywords{Urgent Computing, Edge Computing, Decentralized Computing, Resource Management}


\maketitle

\section{Introduction}\label{sec:introduction}
 
\UC (UC) \cite{boukhanovsky2018urgent, leong2015towards} refers to a class of computing that is specifically designed to support time-critical simulations and computations for urgent events. These events can include natural disasters, crises, or any other situations where timely decisions need to be made based on simulated predictions. The goal of \UC is to provide the necessary computational power and resources to complete these simulations within strict deadlines. 
The importance of \UC lies in its ability to generate timely and accurate results that can aid decision-makers in mitigating financial losses, managing affected areas, and reducing casualties. By simulating the onset and progression of urgent events, authorities can gain valuable insights into the potential impact and make informed decisions to respond effectively. 
\UC does not directly deal with real-time computing or crisis/disaster management. While real-time computing focuses on immediate responsiveness and time-critical processing, \UC emphasizes the simulation and prediction aspect for urgent events. 
However, in crisis and disaster management, a broader range of computational tasks related to managing and mitigating the effects of crises has to be taken into account, including \UC in specific scenarios without being limited to it. 
The canonical interpretation of \UC may result too context-specific, limiting the identification of urgent use cases and hindering the general application of \UC. Therefore, there is a need for more comprehensive and refined approaches to urgent scenarios. 
In fact, the contemporary era is characterized by scenarios where time is of the essence—natural disasters, cyber threats, health crises, and environmental emergencies necessitate rapid and accurate decision-making. 

As a response to the limitations observed in traditional \UC, this paper proposes \textbf{Urgent Edge Computing} (UEC) as an innovative approach. Urgent Edge Computing is positioned as a novel paradigm, acknowledging the limitations of traditional \UC and proposing a dynamic, decentralized, and highly responsive alternative.
\UEC is designed to address the stringent requirements of time-sensitive applications in heterogeneous, distributed edge environments encompassing different kinds of resources accessible using different means, potentially owned by distinct organizations. 
This paper presents and provides a foundational overview of \UEC, outlining its core principles and its potential significance in responding to critical situations promptly and effectively.
%

While Urgent Edge Computing and crisis/disaster computing share the common goal of addressing critical situations, they differ in their focus, scope, and implementation.
Urgent Edge Computing primarily focuses on real-time processing and immediate computations within strict time constraints. The emphasis is on providing rapid and accurate results to support timely decision-making during urgent events. Crisis/Disaster Computing encompasses a broader temporal scope, including pre-event preparedness, response during the crisis, and post-crisis recovery and analysis. 
Urgent Edge Computing places emphasis on processing data near the devices or sensors that generate the urgent data, reducing latency. The geographical focus is on the edge of the network.
Crisis/Disaster Computing involves a wider geographical perspective, addressing crises and disasters at various scales, from local incidents to regional or global events. 

\begin{figure}
    \centering
    \includegraphics[width=1\linewidth]{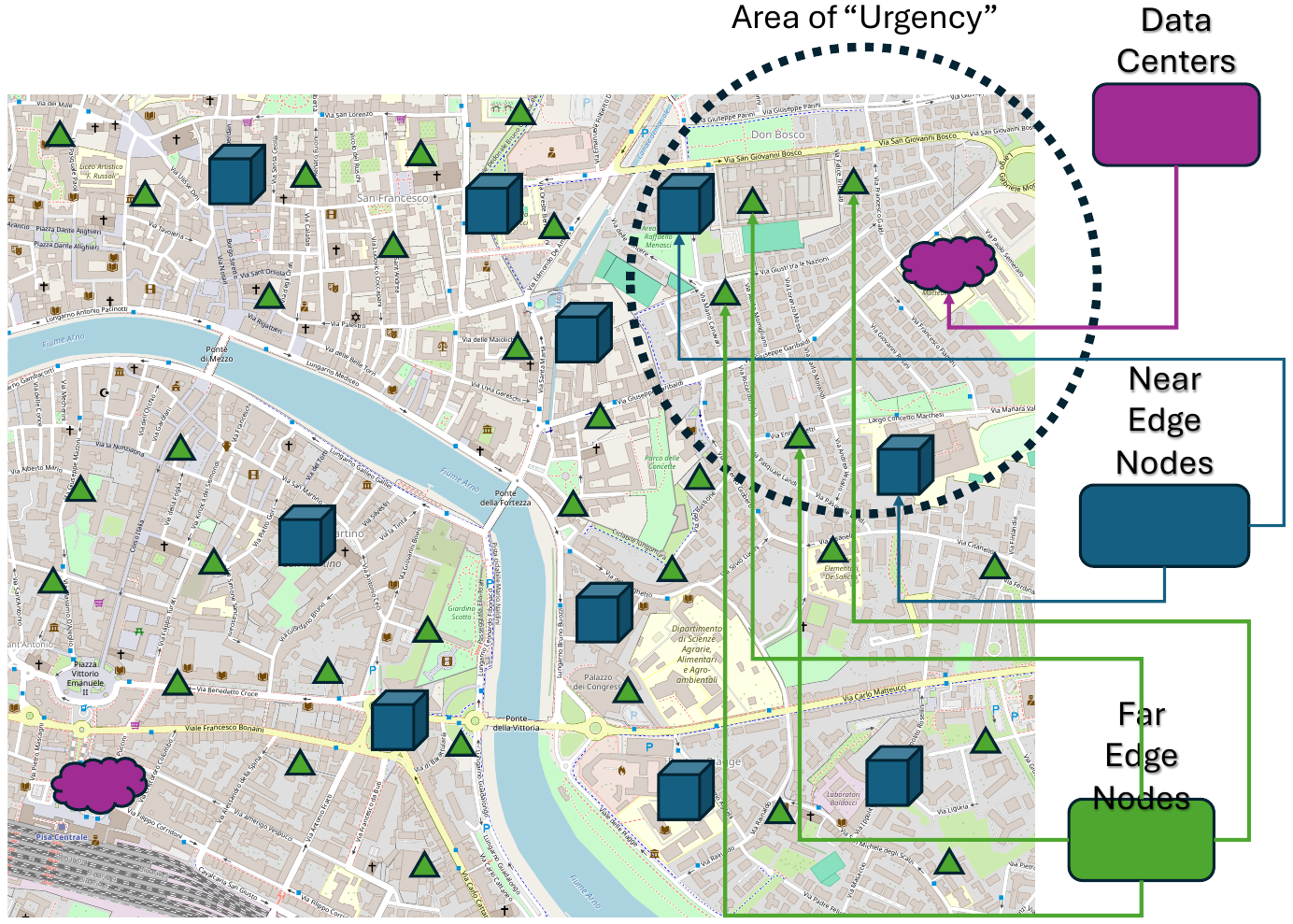}
    \caption{Resources of \UEC}
    \label{fig:UrgentResources}
\end{figure}

Urgent Edge Computing relies on a complex, heterogeneous distributed computing infrastructure, including edge computing resources and portable modular datacenters, to optimize network resources and reduce latency for immediate computations. Crisis/Disaster Computing involves a diverse set of computing resources, which may include cloud computing, high-performance computing clusters, and centralized data centers. The infrastructure is designed to handle large-scale data processing and analytics for comprehensive crisis management.
%
%
\UEC is envisioned to seamlessly support real-time simulation and processing of urgent events using complex, heterogeneous distributed computing resources infrastructures encompassing a heterogeneous set of resources, including those that are available at the edge of the network or can be brought there (e.g., Portable Modular Datacenters). Figure~\ref{fig:UrgentResources} shows a possible example of environment in which \UEC could be applied.
This approach enables immediate computations of various kinds within strict time constraints to support timely decision-making. Urgent edge computing focuses on processing urgent data near the devices or sensors that generate such data, reducing latency and optimizing network resources. 
This allows for fast and accurate results required to address critical situations such as emergencies, natural disasters, or urgent events, enabling relevant authorities to make real-time informed decisions to mitigate losses, manage affected areas, and reduce casualties.

\section{From Urgent Computing to Urgent Edge Computing}\label{sec:headings}

In the context of high-performance computing, \UC stands as a paradigmatic approach that has evolved over the years, providing a way to respond swiftly and effectively to critical situations. This section briefly discusses the history, development, and recent relevance of \UC, describing the context that motivated our proposal for Urgent Edge Computing.
The origin of \UC can be associated with the need for immediate access to powerful computing resources during time-sensitive scenarios. 
This paradigm has its roots in the area of supercomputing, where the imperative to process vast amounts of data became increasingly important. 
Historically, supercomputers have been fundamental in domains such as weather prediction, disaster response, and healthcare, where timely decisions are of paramount importance.
Over the years, \UC has matured and developed alongside advancements in supercomputing technology. The paradigm gained momentum recently. Notably, \UC played a crucial role during the COVID-19 pandemic, where rapid data processing and modeling were essential for understanding virus spread, developing treatment strategies, and optimizing resource allocation in overwhelmed healthcare systems \cite{lopez2021lessons}.

The COVID-19 pandemic highlighted the importance of \UC in supporting institutions during global health crises. As the virus spread rapidly, urgent computations supported epidemiological modeling, drug discovery, and vaccine development. 
Supercomputers worldwide were enlisted to process vast datasets, simulate virus behavior, and conduct complex analyses in record time.
One of the remarkable applications of \UC during the pandemic was evident 
in mapping the virus's genomic code, facilitating the development of targeted therapeutics and vaccines.
The urgency of the situation led to the creation of new approaches and computational methods designed for an immediate response. Job queues were made more efficient, and priority access to supercomputing resources ensured urgent computations related to the pandemic were handled first. The pandemic sparked innovation in \UC, motivating researchers to explore new approaches and technologies to make computational processes faster and more effective.

\subsection{The Transition to \UEC}

As \UC found its place in crucial situations having the need to access massive computational capacity, the evolving technological environment presented both new opportunities and prospects. 
The emergence of Edge Computing and the Internet of Things (IoT) indicated a shift towards decentralized processing, emphasizing the requirement for a more nimble and responsive approach \cite{satyanarayanan2017emergence}. 
This evolution set the stage for the development of \UEC.
\UEC emerges not as a replacement for \UC but as one variation that borrows some of its main concepts and aligns with decentralized demands in the pattern of data and resource access. 

The limitations of traditional \UC, particularly in the context of heterogeneous Edge environments, necessitate a paradigm that can perform computation closer to the data source. 
\UEC departs from the centralized supercomputers approach, but instead distributes decision-making, processing and storage across a network of edge devices.
%
%
%
Thus, even if \UC is a paradigm that aims to provide timely and accurate solutions for critical situations that require immediate intervention, in some scenarios, such as disaster response, environmental monitoring, or smart city management, \UC alone may not be sufficient to meet the stringent requirements of latency, reliability, and scalability. 
For example, in disaster response, edge devices such as drones, sensors, and cameras can collect and process data from the affected area and communicate with each other and with the cloud servers to coordinate rescue operations. Cloud servers that in the context of \UEC are not necessarily remote but can be brought in the area by rescuers and institutions.

In environmental monitoring, edge devices such as satellites, buoys, and weather stations can monitor the changes in the environment and alert the authorities and the public in case of emergencies. 
In smart city management, edge devices such as traffic lights, parking meters, and smart meters can optimize the use of resources and improve the quality of life of the citizens~\cite{cao2019analytics,bacciu2021teaching}. 

This paper introduces \UEC by identifying its key \textit{objectives }(Section~\ref{sec:objetives}) from a algorithmical and methodological standpoint. The paper then identifies some key \textit{technological enablers} that contribute to the development of such objectives (Section~\ref{sec:enablers}). All those enablers are placed and contextualised in a comprehensive \textit{conceptual architecture} for the \UEC (Section~\ref{sec:architecture}). To give to the investigation conducted by the paper a concrete perspective and motivate its relevance, three possible \textit{application scenarios} are presented (Section~\ref{sec:applications}). Finally, conclusions are drawn (Section~\ref{sec:conclusions}).

%
%
%
%

\section{Objectives of the \UEC}\label{sec:objetives}

\UEC applies the principles of \UC at the edge supporting rescuers in their activities while minimizing disruption to running applications. 
\UEC can be functional to its objectives only if it is able to enable the existing infrastructures to provide a certain set of features.

\paragraph{Resource Management} \UEC involves allocating a portion of computational resources specifically for urgent tasks. This can be achieved by implementing resource management mechanisms that prioritize urgent computations over regular applications. By reserving a dedicated share of resources, such as CPU cycles, memory, or network bandwidth, urgent tasks can be given higher priority and guaranteed access to the necessary resources. However, to foster the acceptance of this novel paradigm and avoid having a negative impact on the services currently served by the local infrastructure, this must happen by causing minimal disruption to the resident applications.

\paragraph{Dynamic Resource Provisioning} In urgent situations, resource demands can vary rapidly. To accommodate this, the computing environment should have the ability to dynamically provision additional resources when urgent tasks require more computational power. This can be achieved through techniques such as auto-scaling or load balancing, where resources are allocated on-demand to meet the \UC needs without disrupting existing applications. This is made even more complex when in an attempt to support emergency operations, rescuers bring with them, and pretend to integrate, their own resources into the edge infrastructure.

\paragraph{Quality of Service Differentiation} \UEC should employ mechanisms to differentiate the quality of service provided to the different types of urgent tasks, that may be generated by different applications. This can involve setting strict performance guarantees, such as minimum response times or maximum latency. By ensuring that urgent tasks receive the necessary computational resources and meet their performance requirements, the activities of rescuers need to be supported effectively, possibly without compromising (or minimizing the impact on) the performance of other applications running at the edge.

\paragraph{Priority Scheduling} Implementing priority-based scheduling algorithms can ensure that urgent tasks receive preferential treatment in resource allocation and task execution. Urgent tasks should be scheduled ahead of lower-priority tasks, allowing rescuers to access the computational resources promptly and perform their critical activities without unnecessary delays. This can be achieved by incorporating priority levels, task queues, or preemption mechanisms into the edge computing environment.

\paragraph{Collaboration and Communication} \UEC should facilitate seamless collaboration and communication among rescuers and relevant authorities. This can be achieved by leveraging edge-to-edge and edge-to-cloud communication frameworks, enabling real-time data exchange and coordination between rescuers and the central command center. Such communication channels can ensure that urgent tasks are executed efficiently while maintaining coordination and synchronization among all stakeholders involved. Depending on the actual status of the existing communication and computing infrastructure, in place of the Cloud, it could be envisioned the adoption of mobile datacenters \citemissing.

\paragraph{Self-* properties} 

In urgent scenarios, an \UEC system must possess several self-* properties \cite{babaoglu2004self}. It needs to be self-aware, capable of monitoring its state and environment, and self-learning to improve performance over time. This self-awareness underpins self-configuration and self-optimization, enabling the system to adapt to changing conditions and optimize resources autonomously. Self-awareness also facilitates self-healing and protection, allowing the system to recover from failures and defend against security threats. These properties rely on self-coordination for efficient interaction and decision-making. Additionally, in critical scenarios, self-sustainability is crucial for the system to maintain functionality over an extended period.

\smallskip

By lavishing effort in achieving these objectives, \UEC can effectively reserve and allocate computational resources at the edge, supporting rescuers in their activities without disrupting (or with minimal disruption on) running applications. It can enable the efficient execution of urgent tasks, enhance the responsiveness of rescue operations, and ultimately contribute to more effective crisis management and disaster response. 
The need for this paradigm switch is motivated by the need for urgency in critical scenarios, which demand a computing paradigm that can swiftly adapt, process, and communicate information by encompassing a large set of resources, also including those placed at the edge of the network. 
\UEC strives to bring computation closer to the data (and workload) source while minimizing delays linked to traditional requests and reservations without waiting for unattainable resources.
While these objectives are key for \UEC, most of them are also relevant in the broader context of edge computing. Research is quite active in all these aspects, from priority scheduling to self-* and the support of self-adaptation for QoS provision. Research on \UEC can thus be synergic with recent proposals and advances in the area of edge computing in 6G and ultra-low latency solutions.

\section{Key Enablers of \UEC}\label{sec:enablers}

\UEC is characterized by several key features that differentiate it from traditional computing paradigms and make it particularly suitable for time-critical scenarios. These characteristics enable the processing and analysis of data at the network edge, ensuring ultra-low latency, decentralized decision-making, and self-organization.

\subsection{Ultra-Low Latency Solutions}

\textbf{Scientific Perspective:} Addressing ultra-low latency in \UEC involves interdisciplinary research in several domains. In networking, studies focus on minimizing transmission delays, by adopting the most appropriate routing algorithms, and exploring low-latency communication methodologies and technologies. Moreover, advancements in the way applications are compiled, packaged and deployed on the hardware devices, such as specialized virtual machines (e.g., Unikernels\cite{el2023unikernels}, that are specialized, single-address-space machine images constructed by using library operating systems), or lightweight container runtimes, such as CRI-O\footnote{\url{https://cri-o.io/}}, and more in general, any approach or technology that is able contribute to reducing processing times in the context of a largely heterogeneous and dispersed computational platform.

\noindent\textbf{Challenges and Research Areas:} Researchers have to delve into understanding the fundamental limits of latency in communication networks and computing infrastructures, exploring the impact of various technologies and methodologies, as well as developing innovative techniques for efficient data transmission and processing. Furthermore, investigations into the integration of emerging technologies like 6G and beyond, as well as advancements in edge computing platforms, are critical to achieving ultra-low latency in communication and processing for \UEC.

\subsection{Distributed Processing Algorithms}

\textbf{Scientific Perspective:} Developing efficient and scalable distributed processing algorithms is crucial for \UEC. This area of research encompasses algorithm design, parallel and distributed computing, and optimization techniques. Scientific advancements involve exploring algorithms that can effectively utilize the heterogeneous computational resources of edge devices while ensuring minimal latency.

\noindent\textbf{Challenges and Research Areas:} Scientists need to focus on developing algorithms that can adapt to the dynamic nature of edge environments, considering factors such as varying computing capacities, network conditions, and workload distribution. Optimization of load balancing mechanisms, fault tolerance, and parallel processing techniques are ongoing research areas to enhance the efficiency of distributed processing in \UEC.

\subsection{Security Innovations}

\textbf{Scientific Perspective:} Ensuring the security of \UEC involves advancements in cryptography, authentication protocols, and intrusion detection systems. Scientific research explores novel encryption techniques suitable for resource-constrained edge devices, as well as innovative approaches to secure data transmission, processing and storage in dynamic edge environments, including aspects on confidential computing \cite{zhang2018data}.

\noindent\textbf{Challenges and Research Areas:} Research has to focus on developing lightweight encryption algorithms that balance security and computational efficiency. Additionally, advancements in anomaly detection and intrusion prevention mechanisms are critical to safeguarding \UEC systems against potential cyber threats. The integration of machine learning and artificial intelligence in security protocols is also a burgeoning area of research. 
%

\subsection{Intelligent Edge Orchestration}

\textbf{Scientific Perspective:} Resource Orchestration in \UEC requires solutions and approaches able to satisfy the requests of applications when running on a dynamic set of heterogeneous resources, whose deployment is not necessarily known in advance. Such a scenario calls for research in all those areas that could be beneficial in providing approaches enabling an intelligent orchestration at the edge~\cite{ferrucci2024decentralized,altmann2017basmati}. This includes approaches relying on artificial intelligence, machine learning, and optimization algorithms. Scientific efforts aim to develop intelligent algorithms that dynamically manage and allocate resources, prioritize tasks based on urgency, and adapt to changing environmental conditions. 

\noindent\textbf{Challenges and Research Areas:}
Research is called to explore the development of intelligent algorithms capable of real-time decision-making, taking into account factors like resource availability, energy efficiency, and workload distribution. Optimization of orchestration mechanisms, resource management algorithms, and adaptive learning models are areas of focus to enhance the intelligence of \UEC systems~\cite{dazzi2023intelligent}.
Ongoing research in these areas is essential to unlocking the full capabilities of \UEC, making it a resilient and responsive paradigm for real-time applications in distributed edge environments.

\subsection{Management and Efficient Exploitation of Resource Heterogeneity}

\textbf{Scientific Perspective:} Addressing resource heterogeneity in Urgent Edge Computing involves scientific research in resource management, application structure, load balancing, and adaptive algorithms. Scientific efforts focus on developing algorithms that can efficiently allocate and utilize resources across diverse edge devices, accounting for variations in computing power, storage capacity, and network bandwidth.

\noindent\textbf{Challenges and Research Areas:} Researchers delve into understanding the dynamics of resource heterogeneity in edge environments, exploring algorithms that adapt to changing conditions. Load balancing mechanisms, resource prediction models, and adaptive algorithms are ongoing research areas, aiming to optimize the utilization of diverse resources while maintaining low latency and high responsiveness in Urgent Edge Computing systems. It is worth pointing out that when heterogeneous devices and systems are used, it is fundamental to consider integration as a key challenge as well.

\subsection{Distributed Data Management and Analysis}

\textbf{Scientific Perspective:}
Distributed Data Management and Aggregation in the context of \UEC involves scientific research at the intersection of Big-Data, Data management, Edge computing, Stream processing and Distributed systems. Scientists aim to develop models and algorithms that can operate on decentralized computing infrastructures, facilitating data aggregation and analysis to support real-time decision-making without relying on centralized servers~\cite{mordacchini2009challenges}.

\noindent\textbf{Challenges and Research Areas:}
Research is expected to explore the design of lightweight and efficient data exchange and communication models suitable for highly distributed infrastructures with limited computational and storage resources. Collaborative data approaches and algorithms are areas of focus to enable distributed Data management in \UEC environments. Additionally, investigating privacy-preserving techniques to ensure secure and confidential processing of such data on Edge devices is a critical research avenue, as already mentioned above.

\subsection{Self-Organization and Self-Repair}
Urgent edge computing systems possess self-organization and self-repair capabilities, enabling them to adapt to changing conditions and recover from failures autonomously. In dynamic environments or scenarios where devices may join or leave the network frequently, self-organization ensures seamless integration and collaboration among edge devices. These systems can dynamically allocate resources, optimize task distribution, and establish efficient communication channels based on the available devices and their capabilities. This adaptability enables the system to respond to changing conditions and provide efficient processing and decision-making even in unpredictable environments. Moreover, self-repair mechanisms enable the system to detect and recover from failures or disruptions, ensuring continuous operation and minimizing downtime. The ability to autonomously identify and rectify issues enhances system robustness, scalability, and reliability, making urgent edge computing well-suited for critical scenarios where uninterrupted operation is essential.
\section{Conceptual Architecture}\label{sec:architecture}
The conceptual architecture of \UEC consists of five layers, each encompassing a set of functionalities and services. The list of the layer is the following: 
(i) Computing, Network and Storage: Fabrics; (ii) Computing, Network and Storage: Access, Configuration and Interaction Services; (iii) Computing Network and Storage: Deployment and Management Services; (iv) Platform Abstractions and Execution Services; and (v) Application APIs.

\begin{figure}
    \centering
    \includegraphics[width=\linewidth]{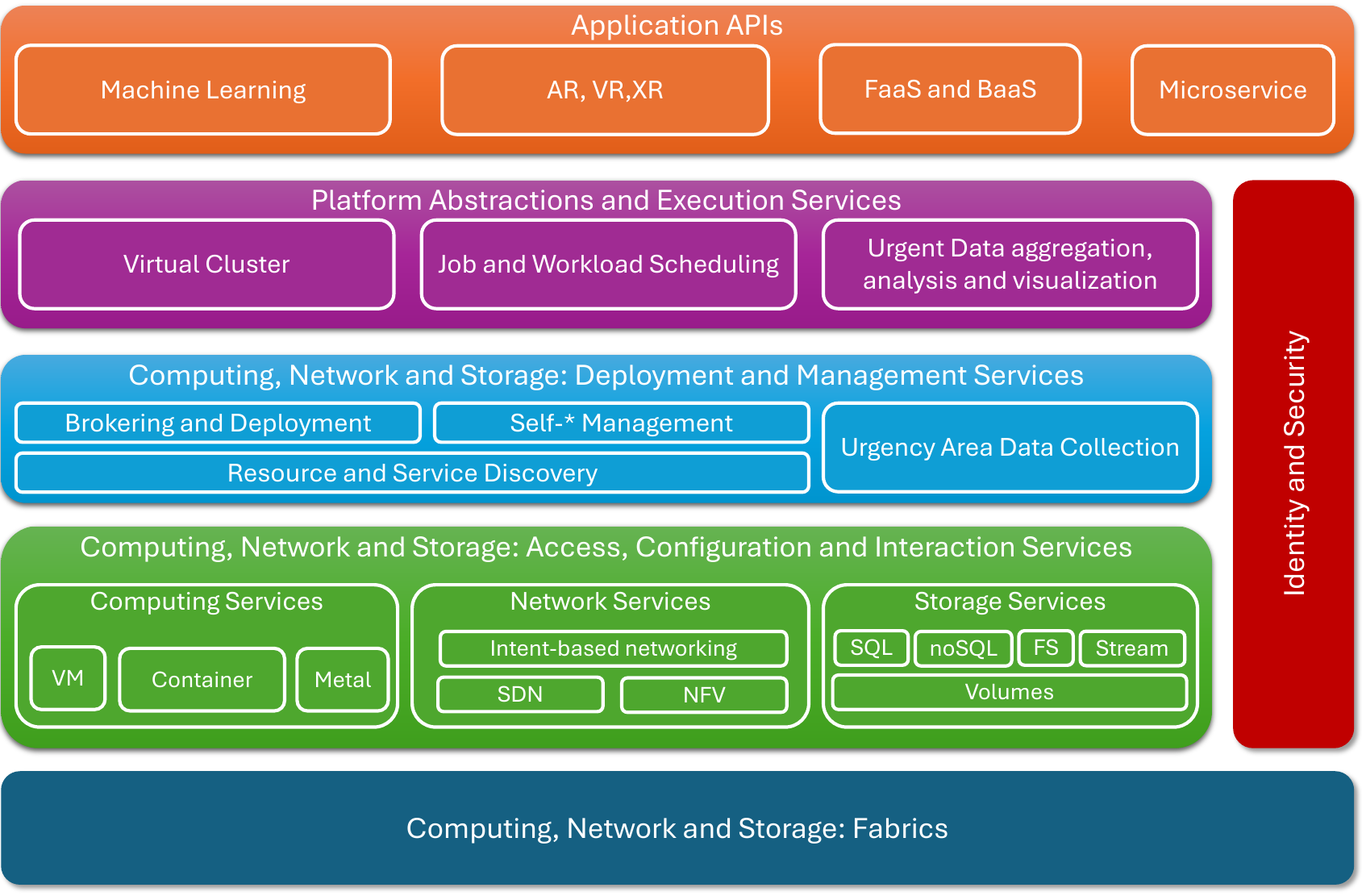}
    \caption{Conceptual Architecture of \UEC}
    \label{fig:arch}
\end{figure}

\subsection{Computing, Network and Storage: Fabrics}
This layer stands at the bottom of the conceptual architecture and represents all the physical resources that will be involved in the computation. It includes any kind of device that is functional to realize the \UEC environment. Such set of entities encompasses both the resources that were pre-existing in the area and those that are brought there to face the needs of \UEC applications.

\subsection{Computing, Network and Storage: Access, Configuration and Interaction Services}
This layer provides all the services that enable access, allow the configuration, and perform the integration with computing, storage, and network services. As such, the layer is organized into three different modules: Computing Services, Network Services, and Storage Services. They include services and tools that enable users or applications to access, configure, and interact with computing resources, network components, and storage systems.

\subsubsection{Computing Services}
This module includes specific mechanisms and approaches for accessing resources and employing different technologies. Specifically, VM, Container, Metal refer to different types of virtualization and resource abstraction. VM stands for Virtual Machine, which emulates a complete computer system (including lightweight virtualization, such as Unikernels). Containers provide a reasonably lightweight and isolated environment for running applications. Metal refers to bare-metal or physical servers without virtualization.

\subsubsection{Network Services}
In this category fall those solutions that provide access to computer networks and enable their configuration. This module is organized into two sublayers. At the bottom stand SDNs and NFVs.
SDN stands for Software-Defined Networking, which separates the control plane from the data plane in network infrastructure, allowing for advanced network management and programmability. NFV stands for Network Functions Virtualization, which virtualizes network functions such as firewalls, routers, and load balancers, running them on standard servers instead of dedicated hardware.

On top of these approaches stands intent-based networking, which employs high-level policies and intentions to automate the configuration and management of network infrastructure~\cite{massa2023declarative}. This is particularly useful in an environment like the one we are considering, where there is a strong need for solutions able to automate networking management to match the complexities of a heterogeneous, dynamic, and multi-tenants environment.

\subsubsection{Storage Services}
Storage services include specific mechanisms and approaches for accessing data and storage systems by adopting different technologies. Also, this module is organized into two sublayers. At the bottom stand the Volumes, and on top of such a layer stands the SQL, noSQL, Filesystems, and data stream layers. Volumes refers to a storage abstraction that represents a logical unit of data storage, often used in the context of virtualization or containerization.

SQL, NoSQL, FS, and Stream are instead different abstraction types over data storage. SQL refers to Structured Query Language, a standard language for managing relational databases. NoSQL refers to non-relational databases that provide flexible data models. FS stands for File System, referring to storage systems that organize data in a hierarchical structure. Stream refers to real-time data streams or event-based data processing.

\subsection{Computing Network and Storage: Deployment and Management Services}
This layer embodies the services and tools needed for deploying, configuring, and managing computing resources, network infrastructure, and storage systems.

It is composed of two main subsystems. One devoted to the collection of data belonging to the area of urgency. This provides the system with the mechanisms and the tools that enable the collection of data from various sources belonging to the Urgency area.
The other subsystem is devoted to the management of the various resources belonging to the system. It is, in turn, organized into two layers. In the bottom part stands the Resource and Service Discovery module that enables the collection of information about resources and services available in the urgency area, including the details on how to connect and access them. On top of it stand two other subsystems: Brokering and Deployment, and Self-* Management.

Brokering and Deployment refer to mediating or coordinating the allocation and deployment of resources or services~\cite{carlini2023smartorc}.
Self-* Management refers to the system's ability as a whole or applications running on top of it to autonomously manage and adapt based on changing conditions or requirements. It is a fundamental feature of large distributed computing platforms~\cite{aldinucci2023proposal}.

\subsection{Platform Abstractions and Execution Services}
Platform Abstractions and Execution Services include the abstraction layers and services that aim at simplifying and automatizing the development, deployment, and execution of applications that target \UEC. This layer includes Virtual Cluster, Job and Workload Scheduling, and Urgent Data aggregation, analysis, and visualization.

\subsection{Virtual Cluster} 
A virtual cluster is a group of virtual or physical resources managed as a unified entity within the context of \UEC. This abstraction layer facilitates resource utilization and scalability, treating the collection of virtualized instances as a single, manageable unit. The virtual cluster allows developers and operators to create, monitor, and maintain a dynamic environment where computing resources are efficiently allocated and scaled based on application requirements. 

\subsection{Job and Workload Scheduling} 
Job and workload scheduling is a critical process involving efficiently allocating computing resources to different tasks or workloads in an optimized manner. This dynamic and sophisticated system aims to support the execution of applications by assigning computing resources based on various factors, including task priorities, constraints, and resource availability. Job and workload scheduling mechanisms utilize algorithms and strategies to ensure that computational tasks are executed effectively, maximizing resource utilization and minimizing latency. By orchestrating the allocation of resources, this process enhances the overall performance, responsiveness, and reliability of applications in \UEC scenarios. The optimization achieved through job and workload scheduling is crucial for meeting strict time constraints and supporting real-time decision-making, which is paramount in addressing urgent events such as emergencies, natural disasters, or time-sensitive situations.

\subsection{Urgent Data aggregation, analysis, and visualization} 
Urgent Data aggregation, analysis, and visualization constitute a pivotal process in \UEC, dedicated to collecting, analyzing, and presenting data in real time. The process begins with the aggregation of data from diverse sources, including sensors, devices, or other data-producing entities at the edge of the network. Subsequently, the collected data undergoes rapid analysis using efficient algorithms to derive meaningful and actionable insights. The final step involves visualization, where the results of the analysis are presented in a clear and comprehensible manner, often through graphical interfaces or dashboards. This real-time data processing and visualization are essential for relevant authorities to make informed decisions promptly, address urgent events with precision, mitigate losses, manage affected areas, and reduce casualties effectively. 

\subsection{Application APIs}
The layers at the top of this conceptual architecture include interfaces or methods designed to ease the development of applications that can benefit from \UEC, encompassing the resources it provides and the services it offers.

Machine Learning (ML) algorithms enable intelligent decision-making by processing and analyzing data in real-time. In Urgent Edge Computing, ML can be employed for predictive analysis, anomaly detection, and automation, enhancing the ability to respond swiftly to urgent events.
AR, VR, and XR technologies enhance visualization and interaction~\cite{theodoropoulos2022cloud,taleb2022toward}. In urgent scenarios, they can provide immersive experiences for real-time data visualization, training simulations, or remote assistance, aiding decision-makers with a more comprehensive understanding of the situation.
FaaS allows developers to execute functions without managing the underlying infrastructure. BaaS provides backend services without the need for extensive backend development. In urgent situations, these services can expedite the deployment of specific functionalities, reducing development time and accelerating response capabilities.
Microservices architecture involves breaking down applications into smaller, independently deployable services. This modularity enhances scalability, agility, and resilience. In Urgent Edge Computing, microservices can facilitate the rapid development, deployment, and scaling of applications, supporting the dynamic nature of urgent events.

\section{Application Scenarios}\label{sec:applications}
In the following, we describe some application scenarios that showcase the advantages offered by \UEC in critical situations.

\subsection{Disaster Response}

In the challenging aftermath of natural disasters like earthquakes, tsunamis, or wildfires, \UEC emerges as a critical solution to overcome traditional computing constraints. Traditional infrastructures often struggle to provide immediate and coordinated responses due to latency issues and centralized processing limitations. \UEC revolutionizes disaster response by enabling the management of edge devices, including drones, sensors, and cameras, strategically deployed in the affected areas. These devices operate in real-time, collecting and processing data locally to gain immediate insights into the extent of damage, identify survivor locations, and assess critical infrastructure conditions.
One of the key advantages of \UEC in disaster response lies in its ability to facilitate seamless communication and coordination among edge devices. These devices collaborate to assess the situation, optimize rescue routes, and deliver essential services like navigation and medical assistance to survivors and responders. By decentralizing decision-making and computations, \UEC ensures that urgent processing occurs closer to the disaster site, significantly reducing response times. This decentralized approach enhances the effectiveness of rescue operations and also contributes to minimizing the overall impact of the disaster on affected communities.

\subsection{Environmental Monitoring}

In the realm of environmental monitoring, Urgent Edge Computing plays a pivotal role in the proactive detection and mitigation of hazards such as air pollution, water contamination, and climate change. The traditional approach to environmental monitoring often involves delayed responses due to centralized data processing and analysis. \UEC transforms this paradigm by strategically deploying edge devices—satellites, buoys, and weather stations—that continuously monitor and analyze environmental changes in real-time.
UEC's adaptive and proactive environmental management capabilities shine through as these edge devices dynamically adjust monitoring parameters based on the analyzed data. For instance, UEC can autonomously modify air quality monitoring parameters or trigger preventive measures in response to imminent threats. The decentralized nature of UEC further enhances the efficiency of environmental monitoring, as localized decision-making ensures that urgent computations occur without relying on centralized systems. This not only accelerates response times but also empowers environmental authorities with accurate and timely information for informed decision-making.

\subsection{Smart City Management} 
\UEC can be exploited in order to improve the quality of life of the citizens and optimize the use of resources in urban areas in case of sudden and/or unpredictable needs. Sensors deployed throughout the city can detect unusual patterns or anomalies in traffic flow, air quality, or noise levels, triggering automated responses or alerts to emergency services for rapid intervention.
Once an area where the need of an intervention has been detected, Edge devices can further coordinate to collect and integrate data from a vast network of heterogenous sources such as traffic lights, parking meters, smart meters, as well as devices available on vehicles, pedestrians, buildings, or utilities. 
\UEC makes it possible for Edge devices to communicate with each other and with cloud servers to easily deploy and coordinate emergency intervention units and allow them to take advantage of localized elements such as traffic control systems, communication units, or energy control devices. 


\subsection{Mass Events Management}
Localized mass events, such as concerts, festivals, rallies, or other crowd events, require real-time, adaptive, and reliable computational support to address critical and time-sensitive needs. \UEC plays a vital role in supporting communication and coordination among various stakeholders involved in event management, including organizers, security personnel, emergency responders, and attendees.
One key aspect is the dynamic and unpredictable nature of mass events, where conditions can rapidly change, and unexpected situations may arise. \UEC enables the rapid analysis of data streams to detect anomalies, identify potential risks, and facilitate timely decision-making. For example, it can help organizers monitor crowd movement patterns and crowd density in real-time to prevent overcrowding or detect security threats. Through real-time data processing and analysis, \UEC systems can optimize resource allocation, traffic flow, and emergency evacuation procedures to ensure the smooth operation of mass events. 

The scalability and flexibility of \UEC systems enable them to adapt to the fast-changing conditions that characterize this kind of scenario. Moreover,  \UEC systems allow data processing and analysis to be performed locally at the edge, minimizing the need to transmit sensitive data to centralized cloud servers. 



\section{Conclusion}\label{sec:conclusions}

The utility and impact of \UC has been demonstrated multiple times, including the recent COVID-19 outbreak. 
\UEC aims to go beyond the traditional concept of the \UC by adopting and extending the underlying infrastructure, including Edge Computing.
We believe that the transition to the \UEC is complex but, at the same time, very necessary. The paper discussed the practical implications of adopting \UEC, such as the potential impact on the infrastructure, resource management, security, and deployment of services on top of a heterogeneous Edge infrastructure.
The inclusion of Edge Computing unlocks the ability to exploit cutting-edge state-of-the-art methodologies and approaches for many modern, decentralized, applications, but also,
most importantly, it makes it possible to reach many more heterogeneous and time-sensitive scenarios in key domains, such as cybersecurity, healthcare, and weather forecasting.

\begin{acks}
This work has been partially funded by Spoke 1 ``FutureHPC \& BigData'' and Spoke 6 ``Multiscale Modelling \& Engineering Applications'' of the Italian Research Center on High-Performance Computing, Big Data and Quantum Computing (ICSC) funded by MUR Missione 4 Componente 2 Investimento 1.4: Potenziamento strutture di ricerca e creazione di ``campioni nazionali di R\&S (M4C2-19 )'' - Next Generation EU (NGEU), and by the NOUS (A catalyst for EuropeaN ClOUd Services in the era of data spaces, high-performance and edge computing) HORIZON-CL4-2023-DATA-01-02 project, G.A. n. 101135927

\end{acks}

\bibliographystyle{ACM-Reference-Format}
\bibliography{main}

\end{document}